\documentclass[12pt,letterpaper,]{article}
\usepackage{rcs}
\usepackage{bm}
\usepackage{amsmath}
\usepackage{amsfonts,helvet}
\usepackage{amssymb}
\usepackage{here}
\usepackage[dviwindo]{graphicx}
\usepackage[below]{placeins}
\usepackage{setspace}
\usepackage{cite}

\RCS $Revision: 1.1 $ 
\RCS $Date: 2004/03/30 02:34:59 $ 
\RCS $Time: $
\DeclareSymbolFont{boldsymbols}{OMS}{cmsy}{b}{n} 
\DeclareSymbolFontAlphabet{\mathbfcal}{boldsymbols}

\begin{document}
\begin{titlepage}

\title{NMTO Wannier-like functions for insulators and metals}
\author{
 Eva Zurek$^a$, Ove Jepsen$^a$, Ole Krogh Andersen$^a$\thanks{
  Author to whom correspondence should be addressed. }
%
%
\\[1ex] \\
$^a$ Max-Planck-Institut f\"ur Festk\"orperforschung, \\
Heisenbergstrasse 1, 70569, Stuttgart, Germany.\\
Fax:(+49) 711-689-1632 \\
Electronic mail: oka@fkf.mpg.de\\[1ex] \\}

\date{\today}
\end{titlepage}

\maketitle
\singlespacing
\newpage

\begin{abstract}
\noindent Within this paper we outline a method able to generate \textit{%
truly minimal} basis sets which describe either a group of bands, a band, or
even just the occupied part of a band accurately. These basis sets are the
so-called \textit{N}MTOs, Muffin Tin Orbitals of order \textit{N}. For an
isolated set of bands, symmetrical orthonormalization of the \textit{N}MTOs
yields a set of Wannier functions which are atom-centered and localized by
construction. They are not necessarily maximally localized, but may be
transformed into those Wannier functions. For bands which overlap others,
Wannier-like functions can be generated. It is shown that \textit{N}MTOs
give a chemical understanding of an extended system. In particular, orbitals
for the $\pi $ and $\sigma $ bands in an insulator, boron nitride, and a
semi-metal, graphite, will be considered. In addition, we illustrate that it
is possible to obtain Wannier-like functions for only the occupied states in
a metallic system by generating \textit{N}MTOs for cesium. Finally, we
visualize the pressure-induced $s\rightarrow d$ transition.\newline
\end{abstract}

\singlespacing

\newpage

\newpage

\newpage \doublespacing

\section{Introduction}

The electronic structure of condensed matter is usually described in terms
of one-electron basis sets. Basis functions used for computation are often
simple, \textit{e.g.} Gaussians or plane waves, but their number is large,
1-2 orders of magnitude larger than the number of electrons to be described.
This is so because the potential in the effective one-electron Schr\"{o}%
dinger equation, arising say in density-functional theory, is rather deep
inside the atoms. The results of this kind of computation therefore require
interpretation and simplification in terms of a small set of intelligible
orbitals. The results of band-structure calculations for crystals, for
instance, are often parametrized in terms of tight-binding models.

The Car-Parinello technique for performing \emph{ab initio}
molecular-dynamics simulations using density-functional theory,
pseudopotentials, plane-wave basis sets, and supercells, \cite{CarPar:1985}
created a need to visualize chemical bonds, and this caused renewed interest
in generating localized Wannier functions for the occupied bands. For a set
of $M$ isolated energy bands, $\varepsilon _{j}\left( \mathbf{k}\right) ,$
the Wannier functions, $w_{m}\left( \mathbf{r-T}\right) ,$ enumerated by $m$ 
$\left( =1,..,M\right) $ and by the $N$ $\left( \rightarrow \infty \right) $
lattice translations, $\mathbf{T,}$ is a set of orthonormal functions which
spans the eigenfunctions,%
\begin{equation}
\Psi _{j\mathbf{k}}\left( \mathbf{r}\right) =N^{-\frac{1}{2}}\sum_{\mathbf{T}%
}\sum_{m=1}^{M}w_{m}\left( \mathbf{r-T}\right) u_{jm}\exp i\mathbf{k\cdot T}
\label{Psik}
\end{equation}%
with eigenvalues $\varepsilon _{j}\left( \mathbf{k}\right) ,$ of the
one-electron Schr\"{o}dinger equation. The Bloch states (\ref{Psik}) are
delocalized, and the set is enumerated by the band index $j$ and the Bloch
vector $\mathbf{k.}$ Wannier functions are not unique, because performing a
unitary transformation, $W_{m,m^{\prime };\mathbf{T-T}^{\prime }},$ of one
set produces another set which also satisfies Eq. (\ref{Psik}), merely with
different $j\mathbf{k}$-dependent phases of the Bloch functions. For
molecules, it had long been recognized that chemical bonds should be
associated with those linear combinations of the occupied molecular orbitals
which are most localized, because those linear combinations are most
invariant to the surroundings. \cite{Boys:1960} For infinite periodic
systems, Mazari and Vanderbilt have developed a useful method for projecting
from the Bloch states a set of so-called maximally localized Wannier
functions. \cite{Marzari:1997, Silvestrelli:1998, Boero:2000}

This paper deals with a different kind of basis set, specifically, minimal
basis sets of localized orbitals, the newly developed \textit{N}MTOs (Muffin
Tin Orbitals of order \textit{N}, also known as $3^{rd}$ generation MTOs. 
\cite{Andersen:2000a, Andersen:2000, Tank:2000, Andersen:2003}) We shall
demonstrate that with \textit{N}MTOs it is possible to generate Wannier
functions directly, instead of via projection from the delocalized Bloch
states. \textit{N}MTOs are constructed from the partial-wave solutions of
Schr\"{o}dingers equation for spherical potential wells (overlapping muffin
tins) and \textit{N}MTO sets are therefore selective in energy. As a
consequence, one can construct an \textit{N}MTO set which picks a specific
set of isolated energy bands. Since \textit{N}MTOs are atom-centered and
localized by construction, they do --after symmetrical orthonormalization--
form a set of localized Wannier functions which, if needed, can be
recombined locally to have maximal localization. The \textit{N}MTO technique
is primarily for generating a localized, minimal basis set with specific
orbital characters, and it can therefore be used also to pick a set of bands
which overlap other bands outside the energy region of interest. The
corresponding \textit{N}MTOs --orthonormalized or not-- we refer to as
Wannier-like.

\textit{N}MTO-generated Wannier functions have so-far been used only in a
few cases to visualize chemical bonding \cite{Andersen:2003, Pavarini:2004}
and more often to construct Hubbard Hamiltonians for strongly correlated 3$d$%
-electron systems. \cite{Muller:1998, Valenti:2001, Pavarini:2001,
Pavarini:2004} In the future, it may be possible to use \textit{N}MTOs for
real-space electronic-structure methods in which the computational efford
increases merely linearly with system size, so-called order-$N$ methods. \cite{Williamson:2001, Williamson:2002}

In this paper we will focus mainly on showing how \textit{N}MTOs may be used
to pick specific states in insulating, semi-metallic and metallic systems,
namely boron, graphite, and cesium. Since \textit{N}MTOs are unfamiliar to
most readers of this issue, and more complicated than plane waves, we start
out by illustrating the main ideas by performing an elementary analytical
calculation of the $\pi $-bonds in the simplest tight-binding (TB) model of
benzene. Then follows a concise summary of the \textit{N}MTO formalism.

\section{\textit{N}MTO basics}

\subsection{Tight-binding calculation of the benzene $\protect\pi $-bond}

The simplest TB model for the six $\pi $-electrons in benzene has six
orthonormal $p_{z}$-orbitals, $\varphi _{1},...,\varphi _{6},$ placed on the
consecutive corners of the hexagon. The hopping integrals over the short
(double) bonds are: $H_{12}=H_{34}=H_{56}=-\left( 1+d\right) ,$ and those
over the long (single) bonds are: $H_{61}=H_{23}=H_{45}=-\left( 1-d\right)$, where $d$ is the dimerization.

In order to calculate the Wannier function for the three occupied bonding
states, it is convenient to \emph{partition} the orbitals into those on the
even $\left( e\right) $- and those on the odd $\left( o\right) $-numbered
sites. The eigenvalue equations are then: $H_{oo}u_{o}+H_{oe}u_{e}=%
\varepsilon 1_{oo}u_{o}$ and $H_{eo}u_{o}+H_{ee}u_{e}=\varepsilon
1_{ee}u_{e},$ in terms of the 3$\times 3$ blocks of the Hamiltonian and the
unit matrices. Solving the last set of equations for the eigenvector for the
even orbitals yields:%
\begin{equation}
u_{e}=\left( \varepsilon -H_{ee}\right) ^{-1}H_{eo}u_{o},  \label{up}
\end{equation}%
and inserting in the first equations results in the eigenvalue equation:%
\begin{equation}
\left[ H_{oo}+H_{oe}\left( \varepsilon 1_{ee}-H_{ee}\right) ^{-1}H_{eo}%
\right] u_{o}=\varepsilon 1_{oo}u_{o},  \label{down}
\end{equation}%
for the (L\"{o}wdin) \emph{downfolded }Hamiltonian.

In the present case where there is no hopping between even or odd sites, $%
H_{ee}=H_{oo}=0.$ Moreover,%
\begin{equation*}
H_{eo}=H_{oe}^{\dagger }=\left( 
\begin{array}{ccc}
-\left( 1+d\right)  & -\left( 1-d\right)  & 0 \\ 
0 & -\left( 1+d\right)  & -\left( 1-d\right)  \\ 
-\left( 1-d\right)  & 0 & -\left( 1+d\right) 
\end{array}%
\right) ,
\end{equation*}%
and%
\begin{equation}
H_{oo}+H_{oe}\left( \varepsilon 1_{ee}-H_{ee}\right) ^{-1}H_{eo}=\frac{1}{%
\varepsilon }\left( 
\begin{array}{ccc}
2\left( 1+d^{2}\right)  & 1-d^{2} & 1-d^{2} \\ 
1-d^{2} & 2\left( 1+d^{2}\right)  & 1-d^{2} \\ 
1-d^{2} & 1-d^{2} & 2\left( 1+d^{2}\right) 
\end{array}%
\right) .  \label{H}
\end{equation}%
The latter, downfolded Hamiltonian is periodic with period $3$ and is
therefore diagonalized by the unitary transformation%
\begin{equation}
U_{ok}=\left( 
\begin{array}{ccc}
\frac{1}{\sqrt{3}} & \frac{-1}{\sqrt{2}} & \frac{-1}{\sqrt{6}} \\ 
\frac{1}{\sqrt{3}} & \frac{1}{\sqrt{2}} & \frac{-1}{\sqrt{6}} \\ 
\frac{1}{\sqrt{3}} & 0 & \frac{2}{\sqrt{6}}%
\end{array}%
\right) ,  \label{U}
\end{equation}%
yielding singly degenerate $a$-states $\left( k=0\right) $ with $\varepsilon
=\pm 2$ and doubly degenerate $e$-states $\left( k=\pm 1\right) $ with $%
\varepsilon =\pm \sqrt{1+3d^{2}}.$ The even components of the eigenvectors
are obtained from equation (\ref{up}): $U_{ek}=\varepsilon ^{-1}H_{eo}U_{ok},
$ and finally we can renormalize: $u=U/\sqrt{2}.$

Having found all six Bloch eigenstates, we need to form three Wannier
functions, that is, three \emph{congruent} linear combinations of the
bonding states. The three \emph{downfolded }$p_{z}$-\emph{orbitals}, $\chi
_{o},$ defined by: $\chi _{o}U_{ok}=\varphi _{o}u_{ok}+\varphi _{e}u_{ek},$
are obviously congruent. Moreover, they are \emph{localized} in the sense
that $\chi _{1}$ vanishes on the other odd sites ($3$ and 5). In the present
case, they are also orthonormal and, hence, Wannier functions.
Left-multiplication with $\left( U_{ok}\right) ^{-1}=U_{ko}$ yields:%
\begin{equation}
\chi _{1}=\frac{1}{\sqrt{2}}\left[ \varphi _{1}+\left( \frac{2}{3}+d\right)
\varphi _{2}+\left( \frac{2}{3}-d\right) \varphi _{6}-\frac{1}{3}\varphi _{4}%
\right] ,  \label{chi}
\end{equation}%
and $\chi _{3}$ and $\chi _{5}$ by cyclic permutation of site indices. Here
and in the following, we work merely to first order in the dimerization, $d.$
The Wannier function in Eq. (\ref{chi}) is essentially the \textit{N}%
MTO. It is atom-centered and, as the dimerization increases, it becomes
lopsided towards site $2,$ \textit{i.e.,} it spills into the short bond. It
breaks the symmetry (when $d\neq 0)$ because it was chosen to vanish on the 
\emph{odd} sites different from its own, and it is not \emph{maximally}
localized (unless $d=0)$. Nevertheless, it is a fairly simple matter to
achieve maximal localization and, hence, to restore the symmetry, by finding
a unitary transformation, $W_{T-T^{\prime }},$ which maximizes \textit{e.g. }%
$\left\langle \left\vert w\right\vert ^{4}\right\rangle \equiv
\sum_{R=1}^{6}\left\vert w\right\vert ^{4}.$ Here, $w_{1}\equiv
\sum_{T}W_{1-T}\chi _{T}$ is the maximally localized Wannier function. In
the present case, $W$ has only one independent matrix element and we find:%
\begin{eqnarray}
w_{1}=\frac{1}{3\sqrt{2}}\Bigg\lbrace \left( \sqrt{3}+1+\frac{\sqrt{3}}{2}d\right)
\left( \varphi _{1}+\varphi _{2}\right)+\left( 1-\sqrt{3}d\right) \left(\varphi _{3}+\varphi _{6}\right) \nonumber \\
 -\left( \sqrt{3}-1-\frac{\sqrt{3}}{2}d\right) \left( \varphi _{4}+\varphi _{5}\right)\Bigg\rbrace ,  \label{w}
\end{eqnarray}%
which is clearly symmetric (bond-centered). From (\ref{chi}): $\left\langle
\left\vert \chi \right\vert ^{4}\right\rangle =\frac{19}{54}+O\left(
d^{2}\right) ,$ and from (\ref{w}): $\left\langle \left\vert w\right\vert
^{4}\right\rangle =\frac{19}{54}+\frac{2\sqrt{3}}{9}d+O\left( d^{3}\right) .$
Hence, for $d=0,$ the atom-centered and the bond-centered Wannier functions
are \emph{both} maximally localized and symmetric.

Now, the \textit{N}MTO set, $\chi _{o}^{\left( N\right) },$ is obtained
without solving the eigenvalue equations, \textit{i.e.,} is \emph{not}
obtained by projection from the Bloch states through multiplication by $%
\left( U_{ko}\right) ^{-1},$ but in a more tricky way: We first define a set
of downfolded, energy-dependent orbitals,%
\begin{equation}
\phi _{o}\left( \varepsilon \right) \equiv \varphi _{o}1_{oo}+\varphi
_{e}\left( \varepsilon 1_{ee}-H_{ee}\right) ^{-1}H_{eo},  \label{phi}
\end{equation}%
which are localized when $\varepsilon $ does not coincide with an eigenvalue
of $H_{ee}.$ Projection onto the even sites yields: $\left\langle \varphi
_{e}\left\vert \hat{H}-\varepsilon \right\vert \phi _{o}\left( \varepsilon
\right) \right\rangle =0_{oo},$ so we realize that the functions of the set $%
\phi _{o}\left( \varepsilon \right) $ are solutions of the impurity problems
specified by the boundary conditions that $\phi _{o}\left( \varepsilon
\right) $ \emph{vanishes} at the \emph{other} odd sites and is normalized to 
$\varphi $ at its own site. Projection onto the odd sites yields:%
\begin{equation}
\left\langle \varphi _{o}\left\vert \hat{H}-\varepsilon \right\vert \phi
_{o}\left( \varepsilon \right) \right\rangle =H_{oo}+H_{oe}\left(
\varepsilon 1_{ee}-H_{ee}\right) ^{-1}H_{eo}-\varepsilon 1_{oo}\equiv
-G_{oo}\left( \varepsilon \right) ^{-1},  \label{G}
\end{equation}%
and comparison with (\ref{down}) shows that, if $\varepsilon $ is an
eigenvalue of the (downfolded) Hamiltonian and $u_{o}$ an eigenvector, then $%
\phi _{o}\left( \varepsilon \right) u_{o}$ is an eigenfunction. $G\left(
\varepsilon \right) $ defined in (\ref{G}) is the resolvent. Finally, we
need to find an energy-independent, $N$th-order approximation to the set $%
\phi _{o}\left( \varepsilon \right) $: We form the set, $\phi _{o}\left(
\varepsilon \right) G_{oo}\left( \varepsilon \right) ,$ of (contracted
Greens) functions and add an analytical function of energy determined in such
a way that the two sets of functions, $\phi _{o}\left( \varepsilon \right)
G_{oo}\left( \varepsilon \right) $ and $\chi _{o}^{\left( N\right)
}G_{oo}\left( \varepsilon \right) ,$ coincide when $\varepsilon $ is on an
energy mesh, $\epsilon _{0},...\epsilon _{N},$ specifying the energy range
of interest. By taking the highest-order finite difference on this mesh, one
obtains:%
\begin{equation}
\chi _{o}^{\left( N\right) }\frac{\Delta ^{N}G_{oo}}{\Delta \left[ 0...N%
\right] }=\frac{\Delta ^{N}\phi _{o}G_{oo}}{\Delta \left[ 0...N\right] },
\label{chiN}
\end{equation}%
which determines the set of (non-orthonormal) \textit{N}MTOs. For $N=1,$ for
instance:%
\begin{equation}
\chi _{o}^{\left( 1\right) }=\left[ \phi _{o}\left( \epsilon _{1}\right)
G_{oo}\left( \epsilon _{1}\right) -\phi _{o}\left( \epsilon _{0}\right)
G_{oo}\left( \epsilon _{0}\right) \right] \left[ G_{oo}\left( \epsilon
_{1}\right) -G_{oo}\left( \epsilon _{0}\right) \right] ^{-1}.  \label{LMTO}
\end{equation}

For the simple benzene model, Eq. (\ref{phi}) yields:%
\begin{equation*}
\phi _{1}\left( \varepsilon \right) =\varphi _{1}-\frac{1+d}{\varepsilon }%
\varphi _{2}-\frac{1-d}{\varepsilon }\varphi _{6},
\end{equation*}%
and $\phi _{3}\left( \varepsilon \right) $ and $\phi _{5}\left( \varepsilon
\right) $ by cyclic permutation of site indices. To order $d,$ the
downfolded Hamiltonian (\ref{H}) is independent of $d,$ and by subtracting $%
\varepsilon $ and inverting, we find:%
\begin{equation*}
G_{oo}\left( \varepsilon \right) =\frac{\varepsilon }{\left( \varepsilon
^{2}-1\right) \left( \varepsilon ^{2}-4\right) }\left( 
\begin{array}{ccc}
\varepsilon ^{2}-3 & 1 & 1 \\ 
1 & \varepsilon ^{2}-3 & 1 \\ 
1 & 1 & \varepsilon ^{2}-3%
\end{array}%
\right) .
\end{equation*}%
Specializing to $N=1,$ the LMTO found from (\ref{LMTO}) is:
\begin{eqnarray*}
\chi _{1}^{\left( 1\right) }=\varphi _{1}-\frac{\epsilon _{0}+\epsilon _{1}}{
\left( \epsilon _{0}\epsilon _{1}+4\right) \left( \epsilon _{0}\epsilon
_{1}+1\right) }\lbrace \left[ \epsilon _{0}\epsilon _{1}+2+\left( \epsilon
_{0}\epsilon _{1}+4\right) d\right] \varphi _{2}+\nonumber \\
\left[ \epsilon_{0}\epsilon _{1}+2-\left( \epsilon _{0}\epsilon _{1}+4\right) d\right]
\varphi _{6}-2\varphi _{4}\rbrace ,
\end{eqnarray*}
and if, with the benefit of hindsight, we choose $\epsilon _{0}=-2$ and $%
\epsilon _{1}=-1,$ we obtain the exact result (\ref{chi}), apart from the
normalization, 1/$\sqrt{2}$. With other choices, the LMTO set is an
approximation to the exact Hilbert space spanned by $\chi _{o}$ or $w_{o},$
as explained in connection with Eq. (\ref{eq2}) below. We leave it to the
reader to convince himself that had we chosen $\epsilon _{0}=2$ and $%
\epsilon _{1}=1,$ we would have obtained the Wannier functions for the
antibonding levels.

By considering a simple TB model, we have thus learned that the \textit{N}%
MTO procedure for constructing a minimal basis set, specifically a set of
localized Wannier functions, consists of the following steps: 1) Downfolding
to a small set of energy-dependent orbitals and 2) a polynomial approximation
of the latter. The resulting \textit{N}MTO set is not orthonormal in
general, but may be symmetrically (L\"{o}wdin) orthonormalized in a third
step. Wannier functions which are maximally localized, and therefore not
symmetry breaking, may be obtained in a fourth step. None of these steps
require knowledge of the extended (Bloch) eigenstates. Although of utmost
importance for applications, steps (3) and (4) are not specific for the \textit{N}MTO
method and fairly standard. As a consequence, they will not be considered
further in this paper.

\subsection{\textit{N}MTOs for real systems}

For real systems, the \textit{N}MTO method constructs a set of atom-centered
local-orbital basis functions which span the solutions of the one-electron
Schr\"{o}dinger equation for a local potential, written as a superposition,\\ $\sum_{R}v_{R}\left( \left\vert \mathbf{r-R}\right\vert \right)$, of
spherically symmetric, short-ranged potential wells, a so-called overlapping
muffin-tin potential. This is done by first solving the radial Schr\"{o}%
dinger (or Dirac) equations numerically to find $\varphi _{Rl}\left(
\varepsilon ,\left\vert \mathbf{r}-\mathbf{R}\right\vert \right) $ for all
angular momenta, $l,$ with non-vanishing phase-shifts, for all potential
wells, $R,$ and for the chosen set of energies, $\varepsilon =\epsilon
_{0},...,\epsilon _{N}$.

The partial-wave channels, $Rlm,$ are partitioned into \emph{active} (odd,
in the benzene example) and \emph{passive} (even). The active ones are those
for which one wants to have orbitals in the basis set. \textit{E.g.} for the
red $\pi $-bands in Figure \ref{fig:gr-pi-all} the active channels are $p_{z}
$ on all the carbon atoms, whereas for the black bands, the active channels
include all nine $s,$ $p,$ and $d$-channels on all the carbon atoms.

For each active channel, $\bar{R}\bar{l}\bar{m},$ a \emph{kinked} partial
wave, $\phi _{\bar{R}\bar{l}\bar{m}}\left( \varepsilon ,\mathbf{r}\right) $
(Eq.\thinspace (\ref{phi}) in TB theory), is now constructed from \emph{all}
the partial waves,\\
 $\varphi _{Rl}\left( \varepsilon ,\left\vert \mathbf{r}-\mathbf{R}\right\vert \right) Y_{lm}\left( \widehat{\mathbf{r-R}}\right)$,
inside the potential-spheres, and from \emph{one} solution, $\psi _{\bar{R}%
\bar{l}\bar{m}}\left( \varepsilon ,\mathbf{r}\right) ,$ of the wave-equation
in the interstitial, a so-called screened spherical wave. The construction
is such that the kinked partial wave is a solution of Schr\"{o}dinger's
equation at energy $\varepsilon $ in all space, except at some hard
screening-spheres --which are concentric with the potential-spheres, but
have no overlap-- where it is allowed to have radial \emph{kinks} in the 
\emph{active} channels. In passive channels, the matching is smooth.

It is now clear that if we can form a linear combination of such kinked
partial waves with the property that all kinks cancel, we have found a
solution of Schr\"{o}dinger's equation with energy $\varepsilon .$ In fact,
this kink-cancellation condition (Eq.s\thinspace (\ref{G}) and (\ref{down})
in TB theory) leads to the classical method of Korringa, Kohn, and Rostoker 
\cite{Kohn:1954} (KKR), but in a screened representation and valid for
overlapping MT potentials to leading order in the potential overlap.

Whereas the screened spherical wave must join \emph{smoothly} onto all \emph{%
passive} partial waves, we can require that it \emph{vanishes} at the hard
spheres for all the \emph{active} channels except the eigenchannel. This
confinement is what makes the screened spherical wave \emph{localized,}
provided that localized solutions exist for the actual potential, energy,
hard spheres, and chosen partition into active and passive channels. Since
the screened spherical wave is required to vanish merely in the \emph{other}
active channels, but not in the eigenchannel, it is an impurity solution for
the hard-sphere solid and is given by Eq.\thinspace (\ref{phi}) in TB theory.

Finally, the set of \textit{N}MTOs is formed as a superposition of the
kinked-partial-wave sets for the energies, $\epsilon _{1},....,\epsilon _{N}:
$%
\begin{equation}
\chi _{\vec{R}\vec{l}\vec{m}}^{\left( N\right) }\left( \mathbf{r}\right)
=\sum_{n=0}^{N}\sum_{\bar{R}\bar{l}\bar{m}}\phi _{\bar{R}\bar{l}\bar{m}%
}\left( \epsilon _{n},\mathbf{r}\right) L_{n;\bar{R}\bar{l}\bar{m},\vec{R}%
\vec{l}\vec{m}}^{\left( N\right) }.  \label{eq1}
\end{equation}%
Note that the size of this \textit{N}MTO basis set is given by the number of
active channels and is independent of the number, $N+1,$ of energy points.
The coefficient matrices, $L_{n}^{\left( N\right) },$ in equation\thinspace (%
\ref{eq1}) are determined by the condition that the set of \textit{N}MTOs
span the solutions, $\Psi _{i}\left( \varepsilon _{i},\mathbf{r}\right) ,$
of Schr\"{o}dinger's equation\emph{\ }with an error%
\begin{eqnarray}
\Psi _{i}^{\left( N\right) }\left( \mathbf{r}\right) -\Psi _{i}\left(
\varepsilon _{i},\mathbf{r}\right)  &=&c^{\left( N\right) }\left(
\varepsilon _{i}-\epsilon _{0}\right) \left( \varepsilon _{i}-\epsilon
_{1}\right) ...\left( \varepsilon _{i}-\epsilon _{N}\right)   \label{eq2} \\
&&+o\left( \left( \varepsilon _{i}-\epsilon _{0}\right) \left( \varepsilon
_{i}-\epsilon _{1}\right) ...\left( \varepsilon _{i}-\epsilon _{N}\right)
\right) .  \notag
\end{eqnarray}%
This condition leads to Eq.\thinspace (\ref{chiN}) and the \textit{N}MTO set
is a polynomial approximation for the Hilbert space of Schr\"{o}dinger
solutions, with $L_{n}^{\left( N\right) }$ being the coefficients in the
corresponding Lagrange interpolation formula. Expression\thinspace (\ref%
{LMTO}) is for \textit{N}=1. An \textit{N}MTO with $N$=0 is a kinked partial
wave, but an \textit{N}MTO with $N>0$ has no kinks, but merely
discontinuities in the $(2N+1)^{st}$ radial derivatives at the hard spheres
for the active channels. The prefactor, $c^{\left( N\right) },$ in
expression (\ref{eq2}) is related to this, \cite{Andersen:2000a} and it
decreases with the size of the set, \textit{i.e.} with the number of active
channels.

A basis set which contains as many orbitals as there are bands to be
described, we shall call \emph{truly minimal. }For an isolated set of bands,
the truly minimal \textit{N}MTO basis converges to the exact Hilbert space
as the energy mesh which spans the range of the bands becomes finer and
finer. Symmetrical orthonormalization of the converged \textit{N}MTO set
therefore yields a set of atom-centered Wannier functions which are
localized by construction. The localization depends on the system \emph{and}
on the choice of downfolding. Had we, for instance for benzene, chosen
instead of the odd sites, sites 1, 2 and 3 as active, the corresponding 
\textit{N}MTO Wannier functions would have been less localized, the Hilbert
space spanned by them would have needed a larger $N$ for convergence, and
the construction from this set of the maximally localized Wannier functions
would have required a larger cluster. Nevertheless, the calculation could
have been done. Similarly, in real systems the choice of active channels and
their hard-sphere radii --their number and main characters being fixed by
the nature of the band to be described-- influences the properties of the
individual \textit{N}MTOs, but not the Hilbert space they converge to.

\section{Results and Discussion}

\label{sec:disc}

\subsection{Graphite: a semi-metal}

\label{subsec:graphite}

The bonding in graphite is understood: within a single graphene sheet the 
\textit{s}, \textit{$p_{x}$} and \textit{$p_{y}$} orbitals on the two carbon
atoms per primitive cell hybridize to form a set of \textit{sp$^{2}$} $%
\sigma $-bonding bands which are occupied, and a set of $\sigma $%
-antibonding bands which are empty. There are two sheets per cell.

The \textit{$p_{z}$} orbitals form a group of bonding and antibonding $\pi $%
-bands which just touch at the Fermi-level, making graphite a semi-metal. In
order to describe the set of $\pi $-bands, we would need to construct two
equivalent \textit{N}MTOs per sheet, each centered on a single carbon atom.
We leave it to the method to shape the orbitals, subject to the
aforementioned boundary conditions for the screened spherical waves. The
energy mesh must be chosen in such a way so that it spans the energy range
of the $\pi $-bands and excludes the energy range where the $\pi $ bands
hybridize with other bands.

\begin{figure}[t!]
\centering\includegraphics[width=0.75\textwidth]{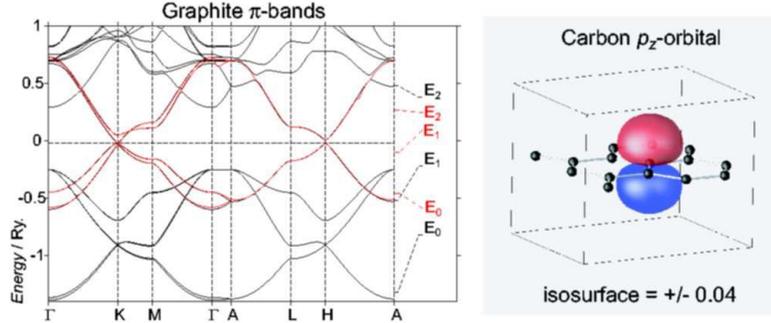}\caption{The band structure of graphite calculated with a full \textit{spd} basis is given in black. The red bands have been calculated with a \textit{$p_{z}$} QMTO (shown on the right) on every carbon atom. The energy meshes used for each calculation are given to the right of the band structure. In all figures, red denotes a positive and blue a negative isosurface value. For the orbital plots, the isosurface values are given in units of $a_B^{-3/2}$, where $a_B$ is the Bohr radius.}
\label{fig:gr-pi-all}
\end{figure}

In Figure \ref{fig:gr-pi-all} the band structure of graphite calculated with
a full $spd$ basis set on each carbon atom is given in black. It is in
excellent agreement with previous calculations. \cite{Ahuja:1995} The red
bands have been calculated with a basis set comprised of one \textit{$p_{z}$}
Quadratic-MTO, or QMTO (\textit{N}=2). The energy meshes used for both
calculations are given to the right of the band structures. The two sets of
bands are almost identical, on the scale of the figure, with the exception
of a small bump at the top of the red bands, where hybridization with other
bands occurs. Thus, it is possible to describe the set of occupied and
unoccupied $\pi $-bands in graphite via just one orbital on every carbon
atom, shown to the right of the band structure. The orbital is localized
because it is not allowed to have any \textit{$p_{z}$} character on any of
the other carbons. It \emph{is} allowed to have other orbital characters,
(E.g. \textit{$s,p_{x},d_{xy}$}), on the other symmetry-equivalent carbons,
but such characters are not visible in the figure.

\begin{figure}[t!]
\centering\includegraphics[width=0.75\textwidth]{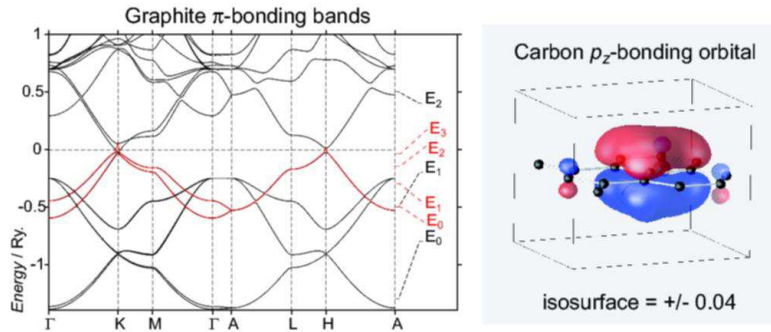}
\caption{As in the previous figure, but the red bands have now been
calculated with a \textit{$p_{z}$} Cubic-MTO (shown on the right) on \textit{%
every second} atom within a single graphene sheet. The energy mesh is chosen
within the occupied part of the $\protect\pi $-band, which is therefore
selected.}
\label{fig:gr-pi-bond}
\end{figure}

\begin{figure}[t!]
\centering\includegraphics[width=0.75\textwidth]{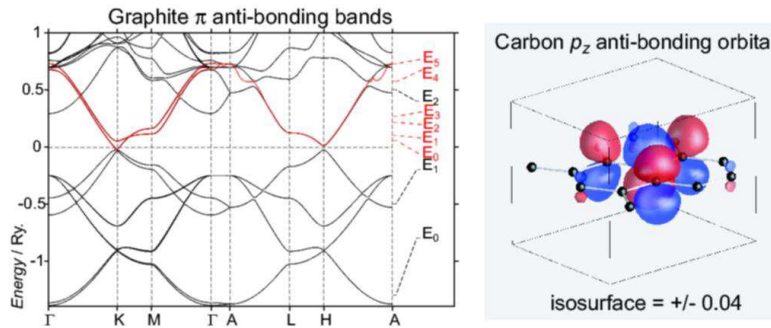}
\caption{As in the previous figure, but the energy mesh is now in the \emph{%
empty} part of the $\protect\pi $-band.}
\label{fig:gr-pi-up}
\end{figure}

It is even possible to generate orbitals for just the occupied or unoccupied 
$\pi $ bands in graphite. In this case we only want to pick half of the
bands, and therefore we need a basis with, say, a \textit{$p_{z}$} orbital
on \textit{every second} carbon atom with all other partial waves being 
\emph{downfolded,} \textit{i.e.} passive. Moreover, an energy mesh spanning
the bonding (anti-bonding) bands must be used in order to obtain the bonding
(anti-bonding) $\pi$ orbital for graphite. Thus, the choice of the energy
mesh determines which set, bonding or anti-bonding, is chosen. In Figures %
\ref{fig:gr-pi-bond} and \ref{fig:gr-pi-up} the orbital on the central
carbon atom is shown, along with the band structures computed with a full 
\textit{spd} basis (in black) and those with the \textit{truly-minimal}
basis set we have just specified (in red). The agreement between the two
sets of bands is excellent, with only minor deviations in the upper regions
of the downfolded band structure of the anti-bonding bands where
hybridization with the next higher bands occurs. Inspection of the $\pi $%
-bonding and anti-bonding orbitals shows that they spread out onto the first
nearest-neighbour carbon atoms (passive), but they are confined not to have
any \textit{$p_{z}$} character on those carbon atoms, e.g. the second
nearest neighbours, where the basis set has orbitals (active partial waves).
The third nearest neighbour atoms also have all partial waves downfolded and
some \textit{$p_{z}$} character may be seen. Clearly, this choice of
orbitals breaks the symmetry; the same Hilbert space would have been
obtained had the orbitals been placed on the other half of the carbon atoms.

\begin{figure}[t!]
\centering\includegraphics[width=0.75\textwidth]{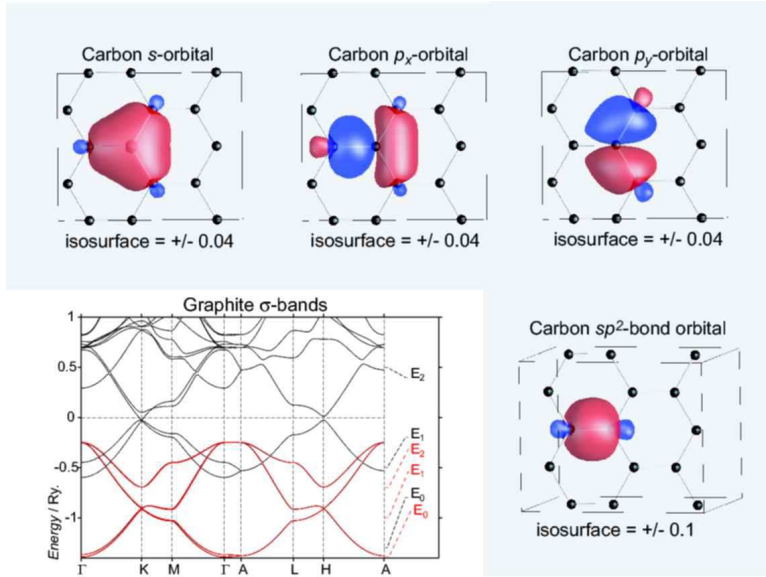}
\caption{As in the previous figures, but the red bands have now been
calculated with an \textit{$s,p_{x}$} and \textit{$p_{y}$} QMTO on \textit{%
every second} carbon atom (shown above the band structure). Also shown is
one of the three congruent \textit{$sp^{2}$}-bond orbitals which arise by
symmetrical orthonormalization of the \textit{$s,p_{x}$} and \textit{$p_{y}$ 
}orbitals.}
\label{fig:gr-pi-sp2}
\end{figure}

It is also possible to describe the \textit{$sp^{2}$}-bonding bands in
graphite by placing an \textit{$s$}, \textit{$p_{x}$} and \textit{$p_{y}$}
orbital on \textit{every second} carbon atom, downfolding all other channels
and using an energy mesh which spans the energy range of the bands of
interest. The band structure obtained with this basis is given in red in
Figure \ref{fig:gr-pi-sp2} and is identical, on the scale of the figure, to
the black bands which have been calculated using a full \textit{spd} basis
set on every carbon atom. The orbitals may spread out onto the nearest
neighbour carbons, however are confined not to have any \textit{$s$}, 
\textit{$p_{x}$} or \textit{$p_{y}$} character on the second nearest
neighbours. Symmetrical orthonormalization of these three orbitals gives the
well known bond orbital, the carbon \textit{$sp^{2}$} \textit{N}MTO, also
shown in the figure.

The above examples show that it is possible to describe a chosen band, or
set of bands, with a truly minimal basis set consisting of one \textit{N}MTO
per band. Moreover, the \textit{N}MTOs obtained from our method are in-line
with an intuitive chemical picture of bonding in the solid state, except
that they may break the symmetry. This arbitrariness originating in the
constraint that the \textit{N}MTOs be atom-centered can be removed by
forming linear combinations of maximally localized Wannier functions. Hence, 
\textit{N}MTOs should be useful not only, for example, as basis sets in
linear scaling methods, but also in gaining a chemical understanding of
periodic systems.

\subsection{Boron Nitride: an insulator}

\label{subsec:bn}

\begin{figure}[h!]
\centering\includegraphics[width=0.75\textwidth]{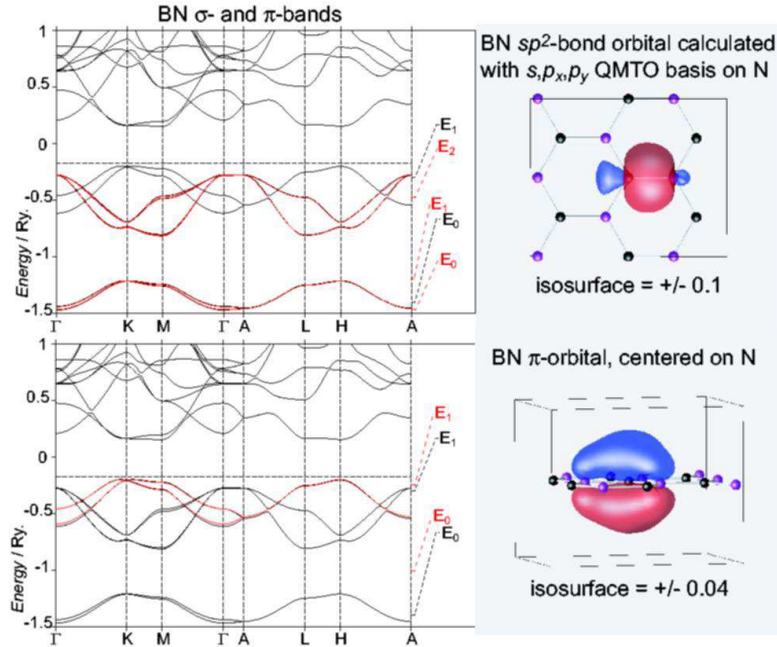}
\caption{The band structure of boron nitride calculated with a full \textit{%
spd} basis is given in black. The red $\protect\sigma $-bonding bands in the
upper panel have been calculated with an \textit{$s,p_{x}$} and \textit{$%
p_{y}$} QMTO on all nitrogen atoms, and the red $\protect\pi $-bonding bands
in the bottom panel with a \textit{$p_{z}$} orbital on all nitrogen atoms.
Also given is one of the three equivalent \textit{$sp^{2}$}-bond orbitals
and the \textit{$p_{z}$} LMTO. Boron atoms are purple, nitrogen black.}
\label{fig:bn-n}
\end{figure}

The bonding in boron nitride is similar to that in graphite: within a single
layer the \textit{s}, \textit{$p_{x}$} and \textit{$p_{y}$} boron and
nitrogen orbitals hybridize to form \textit{sp$^{2}$} $\sigma $-bonding and
-antibonding bands. The alternation, however, makes the system insulating
with a band gap between the bonding and anti-bonding $\pi $-bands. In order
to describe the occupied bands, it is possible to generate either boron or
nitrogen centered \textit{N}MTOs. It can be expected that the electron
density, and hence the orbital at a given isosurface, should have a maximum
closer to the more electronegative element, nitrogen. The method needs to do
less 'work' if the orbitals are placed initially where the electrons are
thought to be. Thus, we first place all the orbitals on nitrogen, and let
the method shape them accordingly. This choice of atom-centered orbitals
corresponds to the extreme ionic limit, a B$^{3+}$N$^{3-}$ configuration.
The bonding $\sigma $ and $\pi $-bands and their respective \emph{N}MTOs are
shown in the top and bottom part of Figure \ref{fig:bn-n}, respectively. The 
\textit{$s,p_{x}$} and \textit{$p_{y}$} \textit{N}MTOs are not shown, since
they are very similar to those obtained for graphite. The full band
structure is in excellent agreement with previous calculations. \cite{Catellani:1987}

\begin{figure}[h!]
\centering\includegraphics[width=0.75\textwidth]{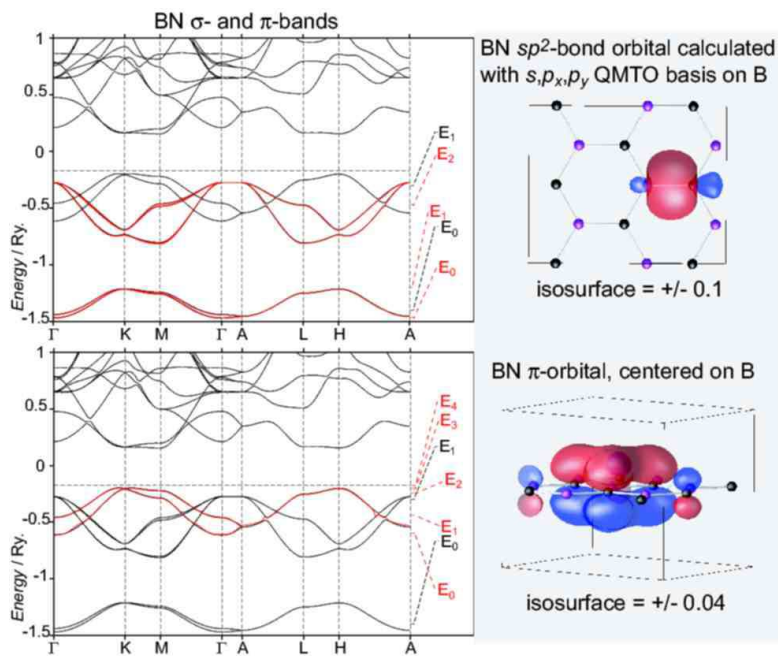}
\caption{As in the previous figure, but with boron-centered $\protect\sigma $
and $\protect\pi$ basis sets.}
\label{fig:bn-b}
\end{figure}

In new materials where the bonding is not well understood, it may be
difficult to decide where the orbitals should be placed. In the following we
will show that \textit{N}MTOs are forgiving: even a bad starting guess can
yield the correct bands and Hilbert space. Placing all the orbitals on the
boron atoms (a B$^{5-}$N$^{5+}$ configuration) yields the bands and orbitals
shown in Figure \ref{fig:bn-b}. The \textit{$sp^{2}$}-bond orbital looks
identical to the one shown in Figure \ref{fig:bn-n}, as it should be when
the energy mesh is converged. For the $\pi $-bands, more energy points are
necessary since the orbital has to spread out from a central boron onto
three neighboring nitrogens. Inspection of the boron (nitrogen) centered $%
\pi $-\textit{N}MTOs makes it plausible that when squared and summed over
all boron (nitrogen) sites, they give the same electron density, with maxima
shifted towards nitrogen.

\subsection{Cesium: a metal}

\label{subsec:cs}

Within this section we will demonstrate that it is even possible to generate
Wannier-like functions that span only the occupied bands of a metal. For
lack of space, we must leave it to the reader to generalize our TB model for
benzene to the corresponding model for the infinite, slightly dimerized
chain, and then to study what happens for vanishing dimerization.

We shall first look at the convergence of the orbital with respect to the
size of the supercell used. Whereas we can only hope to reproduce the
long-ranged Friedel oscillations for supercells so large that the facets of
the folded-in Brillouin zone resemble those of the Fermi surface, much
smaller cells turn out to reproduce the rough shape of the occupied orbital.
This is a manifestation of what Walter Kohn named the "nearsightedness" of
the electronic structure of matter. \cite{Kohn:1996} We shall specifically
consider cesium because of the interesting chemical aspect that its $s$%
-electron transforms into a $d$-electron under hydrostatic pressure.

\begin{figure}[t!]
\centering\includegraphics[width=0.75\textwidth]{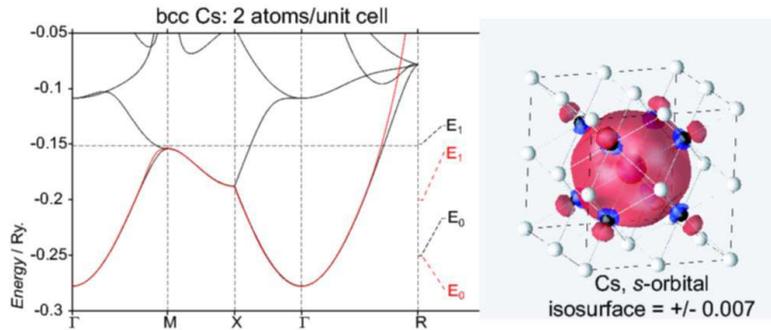}
\caption{Band structure of bcc cesium folded into the 2 atom/cell (CsCl)
simple cubic Brillouin zone. The black bands were calculated with a basis
containing the $s$ and $d$ LMTOs on all atoms. The red bands were calculated
with an \textit{$s$} LMTO on \textit{every second} cesium atom, which is
also shown. The white atoms have an $s$ orbital placed on them (active),
whereas on the black atoms all partial waves are downfolded (passive).}
\label{fig:csI-2}
\end{figure}

The band structure of cesium at ambient conditions (Cs-I) calculated with a
full \textit{sd} basis set on every atom is given in Figure \ref{fig:csI-2}
in black. Superimposed on it in red is the band calculated with one \textit{s%
} orbital on \textit{every second} cesium atom, obtained specifically by
breaking the symmetry and treating Cs metal as CsCl-structured Cs$^{+}$Cs$%
^{-}.$ In most regions of the Brillouin zone the agreement between the two
is good, except it is obvious that with this supercell it is not possible to
describe the occupied part of the upper band along the $\Gamma $X-line.
Nonetheless, the orbital can be plotted and is shown in Figure \ref%
{fig:csI-2}.

\begin{figure}[t!]
\centering\includegraphics[width=0.75\textwidth]{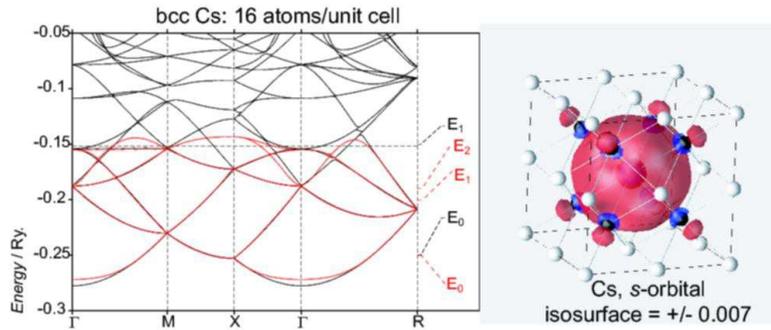}
\caption{As in the previous figure, but for a 16-atom supercell.}
\label{fig:csI-16}
\end{figure}

The result obtained by doubling the cell in all three directions is shown in
Figure \ref{fig:csI-16}. Now the occupied band structure has improved
considerably, but is not yet perfect. The body of the orbital obtained from
this calculation is, however, very similar to the one generated from the
CsCl supercell. The long-ranged tail we do not monitor with the contour
chosen in the figures. Already, this orbital at low isosurfaces shows the
onset of $sd$-hybridization. At high isosurfaces (not shown), the orbital is
completely $s$-like. The hybridization is evident in the \textit{$d_{z^{2}}$}
character found on the nearest-neighbour atoms which have all partial waves
downfolded, and therefore can posses any orbital character. It is a result
of the fact that even though the occupied band has primarily \textit{s}
character, near the Fermi level some regions with \textit{$t_{2g}$} and 
\textit{$d_{z^{2}}$} character can be found.

Under pressure, cesium undergoes a variety of interesting structural phase
transitions. At 2.3 GPa Cs-I (bcc cesium) transforms to an fcc phase.\cite{Hall:1964} Until recently, it was believed that Cs-II undergoes an isostructural transition to Cs-III (which is found in a very narrow pressure range between $\sim $4.2 and $\sim $4.3 GPa). However, experiments have shown that Cs-III has a very complicated structure which is orthorhombic (space group $C222_{1}$ with 84 atoms per unit cell).\cite{McMahon:2001} At $\sim $4.3GPa Cs-III transforms to the non-close-packed tetragonal Cs-IV 
\cite{Takemura:1982} which at $\sim $12GPa undergoes a transition to
orthorhombic Cs-V \cite{Schwarz:1998} and finally to the double hexagonal
close packed Cs-VI at $\sim $70GPa. \cite{Takemura:2000} These structural
transitions are beleived to be driven by the pressure-induced $s\rightarrow d
$ valence electron transition. \cite{Sternheimer:1950} Here we will generate 
\textit{N}MTOs for Cs-II and Cs-IV in order to visualize how the valence $s$
orbital in cesium changes with increased pressure.

The occupied bands for Cs-II with $v/v_0=$ 0.6 are still primarily $s$-like,
however a fair amount of $e_g$-character can also be found. Upon increasing
pressure, a Lifshitz transition occurs (change in the topology of the Fermi
surface) as higher lying $t_{2g}$ bands cross the Fermi level. Figure \ref%
{fig:csII} shows the band structure of Cs-II for a 32-atom supercell
compressed past the calculated volume of the Lifshitz transition. The
agreement between the downfolded and full band structures is excellent and
the downfolded bands are even able to describe the aforementioned $t_{2g}$
bands. The orbital at high isosurfaces is clearly no longer $s$-like. It
must be noted that the orbital calculated for $v/v_0=$ 0.6, using the same
supercell and downfolding scheme, yielded a very similar orbital (not
shown), the only difference being that the four lobes evident in Figure \ref%
{fig:csII} are located in the same plane.

\begin{figure}[t!]
\centering\includegraphics[width=0.75\textwidth]{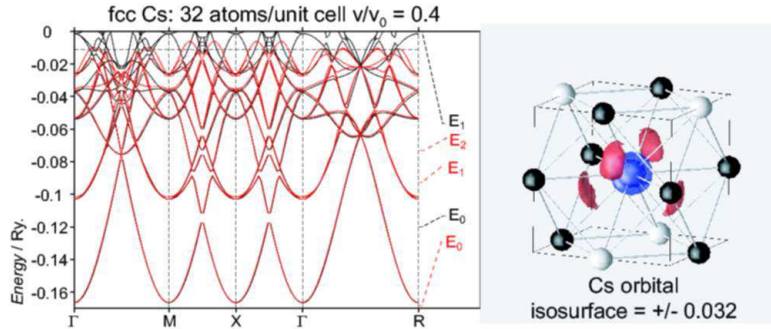}
\caption{As in the previous figure, but for fcc cesium with $v/v_0=$ 0.4,
and a 32-atom supercell.}
\label{fig:csII}
\end{figure}

\begin{figure}[b!]
\centering\includegraphics[width=0.75\textwidth]{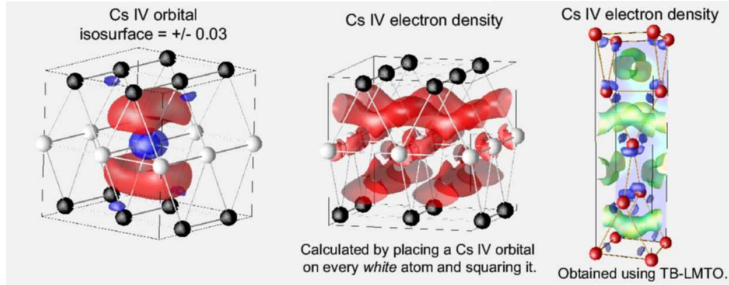}
\caption{The LMTO obtained for Cs-IV ($v/v_0=$ 0.316) with 2-atoms/cell. An $%
s$ orbital was placed on all white atoms (active), whereas the partial waves
on the black atoms were downfolded (passive). Also shown is the charge
density obtained by placing the LMTO on every white atom and squaring it,
along with the charge density obtained from a TB-LMTO calculation. The
latter charge density is colored by the ELF. The isosurface taken is 0.0063 $a_B^{-3}$.}
\label{fig:csIV}
\end{figure}

Cs-IV with each cesium atom having a coordination of 8 is no longer a
close-packed structure. It can be viewed as a stacking of prisms with a
ninety degree rotation from layer to layer in the $c$-direction. The TB-LMTO
calculated charge density (Figure \ref{fig:csIV}) shows maxima in the
interstitial regions, in the center of these prisms. The LMTO obtained by
placing one $s$-orbital on every second cesium atom and downfolding all
other partial waves is given in Figure \ref{fig:csIV}. Clearly this orbital
can be obtained from that shown in Figure \ref{fig:csII} by raising two and
lowering two of the lobes. Placing the Cs-IV LMTO on all of the active sites
and squaring it yields a charge density which is almost identical to that
calculated with TB-LMTO, \cite{Andersen:1984} giving further validation that
our method works. Thus we have shown that $N$MTOs may be used to give a
chemical picture of the pressure induced electronic phase transition in
cesium yielding results which are in-line with those obtained from standard
electronic structure calculations.


\section{Summary and conclusions}

Within this article we have shown that the $3^{rd}$ generation MTO method
can be used to design a basis set of atom-centered localized orbitals, which
span the wave functions in a given energy range. For an isolated set of
bands, arbitrary accuracy may be obtained, with only one basis function per
electron pair (or single electron, in the case of spin-polarized
calculations), simply by increasing the density of the energy mesh which
spans the energy range of the band in question. This method may be applied
to insulating and even metallic systems to generate Wannier-like functions
which are in-line with a chemical understanding of bonding in the
solid-state. It should therefore be a useful analysis tool in, for example,
explaining experimental trends for a given set of compounds via orbital
based arguments; clarifying the bonding in novel or even amorphous
materials; visualizing pressure-driven electronic transitions.

$3^{rd}$ generation MTOs may also be useful in generating \textit{%
truly-minimal} basis sets for order-N methods and in constructing
many-electron wave functions which can be applied to study strongly
correlated systems realistically. In our implementation, \textit{N}MTOs are
generated using the self-consistent potential from an LMTO calculation.
However, our method may be interfaced with the results of any other program,
providing that the potential can be expressed in terms of a superposition of
spherically symmetric potential wells with radial overlaps of up to $\sim $%
60 percent.

\section{Computational Methods}

Graphite has a hexagonal unit cell. The space group is P6$_3$/mmc (194) and
the two basis carbon atoms are located in the $2b$ and $2c$ Wyckoff
positions. The lattice constants, \textit{a} and \textit{c} were taken from
experimental data \cite{graphite} as being 2.4642 {\AA } and 6.7114 {\AA },
respectively. In addition to the atoms on the two crystallographic
positions, it was necessary to insert two interstitial spheres to represent
the charge density in the calculation.

Boron nitride is also found with space group P6$_3$/mmc (194), with the
boron atom located in the $2c$ and the nitrogen atom in the $2d$ Wykoff
positions. The lattice constants, \textit{a} and \textit{c} were taken from
experimental data \cite{bn} as being 2.50399 {\AA } and 6.6612 {\AA },
respectively. It was only necessary to insert one interstitial sphere.

At ambient conditions, cesium crystalizes in the bcc structure. Calculations
were performed using the experimental lattice parameter of 6.048 {\AA }. \cite{bcc:cs} Cs-II is fcc and the bands shown here were calculated for $v/v_0=$ 0.4. Both Cs-I and Cs-II are close-packed so it was not necessary to insert any empty spheres. Cs-IV has the space group I4$_1$/amd with 2 atoms per unit cell \cite{Takemura:1982} and it was necessary to insert four empty spheres.

All of the TB-LMTO calculations \cite{Andersen:1984} were performed using
the Vosko-Wilk-Nusair (VWN) LDA \cite{Vosko:1980} along with the Perdew-Wang
GGA. \cite{Perdew:1986} Scalar relativistic effects were included. For
graphite and boron nitride a basis set consisting of $spd$ LMTOs on the
carbon, boron and nitrogen atoms with $sp$ LMTOs on the empty spheres, was
used. For cesium, the basis consisted of $sd$ LMTOs with $pf$ LMTOs being
downfolded. For graphite and boron nitride, the calculations utilized 1953
irreducible points in the tetrahedron \textit{k}-space integrations,\cite{Blochl:1994} 1661 and 897 points were used in the calculation for cesium in the bcc and fcc structures (one atom per unit cell), respectively, and 693 points were used in the calculation for Cs-IV.

The present version of the \textit{N}MTO program is not self-consistent and
requires the output of the self-consistent potential from the TB-LMTO
program. The downfolded band structures are compared with bands computed
employing a full \textit{N}MTO basis set; \textit{not} with those obtained
using the TB-LMTO program. In all cases, the default values for the
hard-sphere radii, $a_{R}$, were used. Thus, all of the hard spheres were
slightly smaller than touching. However in general, the $a_{R}$ should be
taken as 0.9 times the tabulated covalent, atomic or ionic radii. In the
calculations all partial-waves on the empty spheres were downfolded. The
other downfolding schemes and energy meshes employed for particular
calculations are given in the results and discussion section of the paper.
Note that a mesh employing two (\textit{N}+1=2) energy points yields a
linear-MTO or LMTO, one with three points a quadratic-MTO or QMTO, and so
on. More details about the \textit{N}MTO formalism can be found in \cite%
{Andersen:2000, Andersen:2003, Tank:2000} and references within.


\section*{Acknowledgments}

E.Z. acknowledges financial support from the 
\textquotedblleft International Max-Planck Research School for Advanced
Materials" (IMPRS-AM).


\bibliographystyle{unsrt}
\bibliography{structure,lmto,nmto,apps,cs}


\end{document}